\documentclass[referee]{aa}
\usepackage[varg]{txfonts}
\usepackage{natbib}
\bibpunct{(}{)}{;}{a}{}{,} 
\begin{document}
	
\title{Horizontally Polarized Kink Oscillations Supported by Solar Coronal Loops in an Asymmetric Environment}
	
\author{Mijie Shi\inst{1}
	\and Bo Li\inst{1}
	\and Shengju Yuan\inst{2}}
	
\institute{Shandong Key Laboratory of Optical Astronomy and Solar-Terrestrial Environment, School of Space Science and Physics, Institute of Space Sciences, Shandong University, Weihai, Shandong, 264209, China\\
\email{shimijie@sdu.edu.cn}
	\and Institute of Frontier and Interdisciplinary Science, Shandong University, Qingdao, Shandong, 266237, China}

\titlerunning{Kink Oscillations in Coronal Loops with an Asymmetric Environment}
\authorrunning{Shi et al}

\date{Received ......... / Accepted .........}
	
\abstract{Kink oscillations are ubiquitously observed in solar coronal loops,
		their understanding being crucial in the contexts of coronal seismology and atmospheric heating.}
		{We study kink modes supported by a straight coronal loop  embeded in an asymmetric environment
		using three-dimensional magnetohydrodynamic (MHD) simulations.} 
		{We implement the asymmetric effect by setting different exterior densities 
		below and above the loop interior,
		and initiate the simulation using a kink-like velocity perturbation perpendicular to the loop plane,
		mimicking the frequently measured horizontally polarized kink modes.}
		{We find that the external velocity fields show fan blade structures propagating in the azimuthal direction
		as a result of the successive excitation of higher azimuthal Fourier modes.
		Resonant absorption and phase mixing can still occur despite an asymmetric environment,
		leading to the development of small scales at loop boundaries.
		These small scales nonetheless develop asymmetrically at the upper and lower boundaries
		due to the different gradients of the Alfv\'en speed.}
		{These findings enrich our understanding of kink modes 
		in coronal loops embedded within an asymmetric environment,
		providing insights helpful for future high-resolution observations.} 
	
	 
\maketitle 

\section{Introduction}
\label{S-Introduction}
Magnetohydrodynamic (MHD) waves abound in the solar atmosphere
\citep[see reviews by, e.g.,][]{2020SSRv..216..136L,2020SSRv..216..140V,2021SSRv..217...76B,2021SSRv..217...73N,2021SSRv..217...34W}.
In particular, kink waves supported by the structured corona 
have attracted considerable attention since
their first imaging detection
\citep{1999ApJ...520..880A,1999Sci...285..862N}.
Further observations show that kink oscillations in coronal loops may be
decaying \citep[e.g.,][]{2015A&A...577A...4Z,2016A&A...585A.137G} or 
decayless \citep{2012ApJ...759..144T,2012ApJ...751L..27W,2013A&A...552A..57N,2015A&A...583A.136A}.
Ubiquitous propagating kink modes were also detected
in the outer corona
\citep{2007Sci...317.1192T,2008ApJ...676L..73V}.
Studies of kink modes are often placed in the context of 
either coronal heating or coronal magnetic field diagnosis
\citep[e.g.,][]{2001A&A...372L..53N,2020Sci...369..694Y}.
The two contexts are not mutually exclusive.
As an example, \cite{2020A&A...638A..89G}
connected the evolution of kink waves with the thermal evolution of their host loops.

A thorough understanding of kink modes is crucial for their applications.
In classical theories where a straight coronal loop and
a piecewise constant loop interior and exterior are assumed,
kink modes are a collective eigenmode with azimuthal wavenumber $q=1$
\citep[e.g.,][]{1983SoPh...88..179E}.
When a continuous transition layer connects the loop interior and exterior,
kink modes are subject to resonant absorption \citep{2011SSRv..158..289G}.
During this process, the coordinated wave energy is transferred to
localized Alfv\'enic motions, which subsequently phase-mix
to generate small scales.
Along this line of thinking, a variety of modeling works have been performed 
to explore either the properties of kink modes or their heating effects
\citep[e.g.,][]{2015ApJ...809...72A,2016A&A...595A..81M,2017A&A...604A.130K,
2019ApJ...870...55G,2020ApJ...904..116G,2021ApJ...908..233S,2021ApJ...922...60S,2022ApJ...941...85D}.

In reality, however, the environment of coronal loops is unlikely to be uniform.
Take the much-studied horizontally polarized kink modes \citep{2008A&A...489.1307W}.
The upper and lower ambient coronae of the loop apex  may be asymmetric due to density stratification.
While the properties of MHD waves in an asymmetric slab configuration have been examined \citep[e.g.,][]{2017SoPh..292...35A,2018ApJ...853..136K,2022ApJ...940..157C},
there exists no dedicated study on collective waves supported by an asymmetric tube.
This work examines the property of kink modes supported by a straight coronal loop 
in an asymmetric environment using 3D MHD simulations.
New features are revealed in comparison with the case where a symmetric ambient corona is implemented.
Section~\ref{S-setup} provides the numerical setup,
followed by the simulation results in Section~\ref{Results}.
Section~\ref{S-summary} is the summary.

\section{Numerical setup}
\label{S-setup} 
\begin{figure}    
	\begin{center}
		\resizebox{\hsize}{!}{\includegraphics[width=\textwidth,clip=]{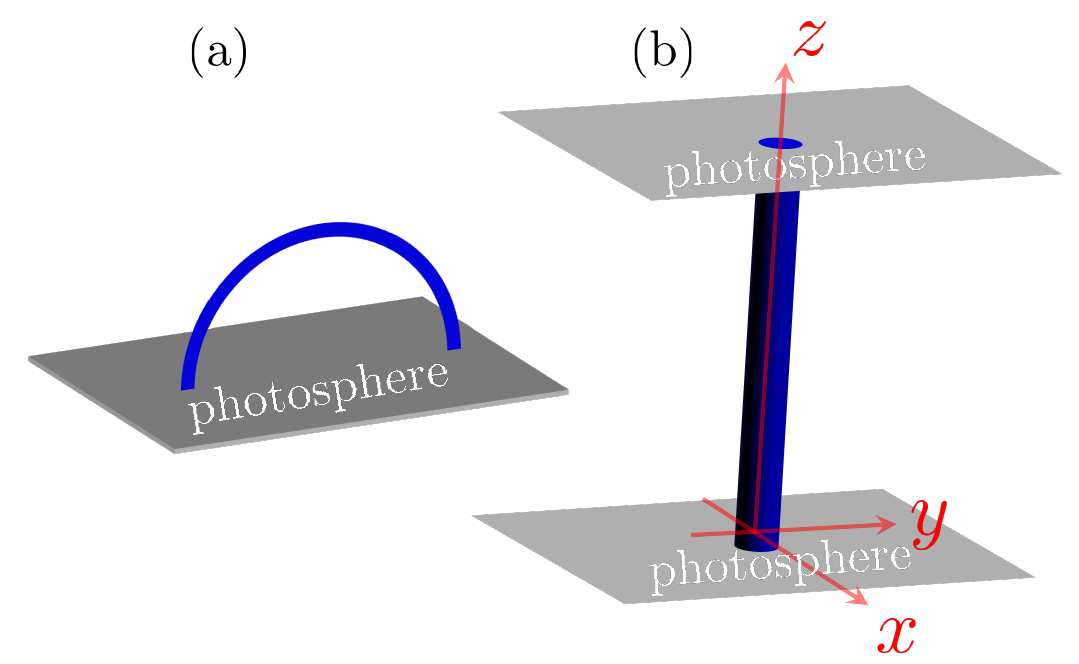}}
		\caption{Illustrations of (a) a curved tube anchored in the photosphere
			and (b) its straightened version used in our configuration.}
		\label{f1}
	\end{center}
\end{figure}

\begin{figure}    
	\begin{center}
		\resizebox{\hsize}{!}{\includegraphics[width=\textwidth,clip=]{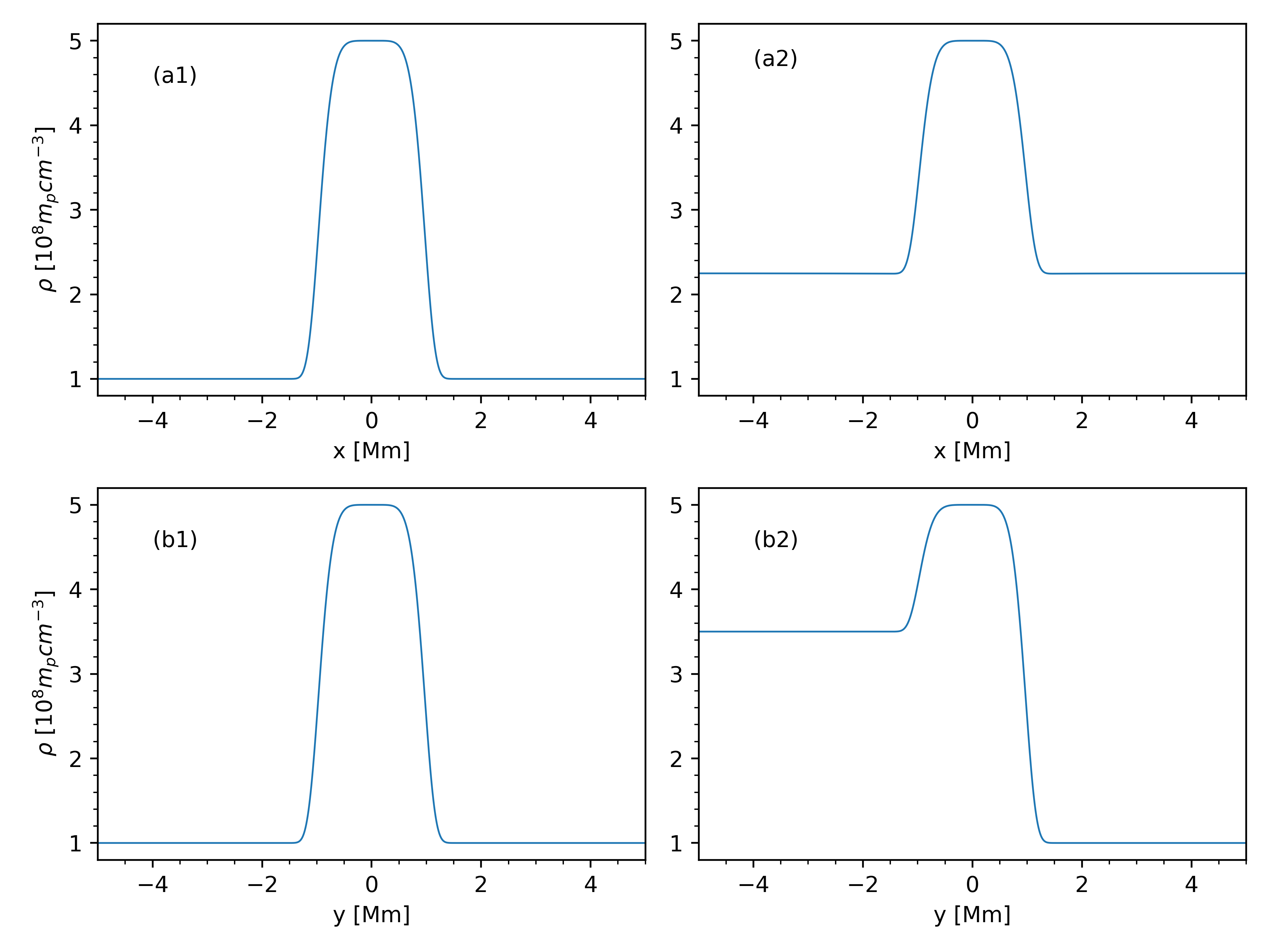}}
		\caption{Equilibrium density profiles along the $x$-axis ($y=0$) and $y$-axis ($x=0$) for the symmetric (left) and 
		asymmetric case (right).}
		\label{f2}
	\end{center}
\end{figure}

Our equilibrium configuration, illustrated in Figure~\ref{f1}(b), 
is a straightened version 
of a curved tube anchored in the photosphere (Figure~\ref{f1}(a)). 
We see coronal loops as field-aligned magnetic tubes. 
The equilibrium magnetic field is $z$-directed, 
and all equilibrium quantities are $z$-independent.
The equilibrium density is prescribed by
\begin{eqnarray}
	\rho(x,y) = \rho_{\rm e}(\theta)+\left[\rho_{\rm i} - \rho_{\rm e}(\theta)\right]f(r),
\end{eqnarray}
where $(r,\theta)$ denotes a standard polar coordinate system
in the $x-y$ plane and the azimuthal coordinate  $\theta$ is 
measured counter clockwise from the positive direction of the $x$-axis.
A continuous profile $f(r) ={\rm exp}[-(r/R)^{\alpha}]$ is used to connect 
the internal density $\rho_{\rm i}=5\times10^{8}m_{\rm p}{\rm cm}^{-3}$ ($m_{\rm p}$ is the proton mass)
to some external density $\rho_{\rm e}(\theta)$,
with $\alpha=6$ and the nominal loop radius $R=1$~Mm.
This setup makes the length of the transition layer $l$ normalised to the loop radius as $l/R\approx0.6$,
which is within the range of some seismological estimation based on resonant absorption
\citep[e.g.,][]{2007A&A...463..333A}.
We construct an asymmetric background (hereafter the asymmetric case) by setting 
\begin{eqnarray}
	\rho_{\rm e} (\theta)= \rho_{\rm e1}  + (\rho_{\rm e2} - \rho_{\rm e1})\frac{1-\rm{sin}\theta}{2},
	\label{rho_profile}
\end{eqnarray}
where
$[\rho_{\rm e1}, \rho_{\rm e2}]$=$[1,3.5]\times10^{8}m_{\rm p}{\rm cm}^{-3}$.
Here  $\rho_{\rm e1}$ represents the density at $(x,y)=(0,+\infty$)
and $\rho_{\rm e2}$ at $(x,y) = (0,-\infty$).
Figures~\ref{f2}(a2) and \ref{f2}(b2) show the equilibrium density profiles along
	the $x$-axis (i.e., $y=0$) and $y$-axis (i.e., $x=0$) for this case. 
 One sees that the loop environment is asymmetric along the $y$-direction but symmetric along the $x$-direction.
 The $x$-($y$-) direction is seen as horizontal (vertical) in accordance with the convention 
 in naming the polarization properties of kink motions (see Figure~\ref{f1}(a)).
The magnetic field is set to be uniform ($15$~G).
The interior temperature at $(x,y)=(0,0)$ is $T_{\rm i}=1$~MK.
A uniform thermal pressure is implemented to enforce transverse force balance,
the resulting plasma $\beta$ being 0.0155.
The pertinent Alfv\'en speeds are $[v_{\rm Ai},v_{\rm Ae1},v_{\rm Ae2}]=[1463,3272,1749]~\rm{km~s^{-1}}$.
The loop length is $L=50$~Mm.
A further simulation with the symmetric background 
($\rho_{\rm e1}=\rho_{\rm e2}$=$1\times10^{8}m_{\rm p}{\rm cm}^{-3}$, hereafter the symmetric case)
is conducted for comparison
( Figures~\ref{f2}(a1) and (b1)).

For both cases, the equilibrium is perturbed by a horizontal kick
\begin{eqnarray}
	\label{eq2}
	v_x(x,y,z;t=0)=v_0 {\rm exp}\left(-\frac{r^2}{2\sigma^2}\right){\rm{sin}}\left(\frac{\pi z}{L}\right),
\end{eqnarray} 
where $v_0=10{\rm~km~s^{-1}}$ is the amplitude,
and $\sigma=R$ characterizes the spatial extent.
The $z$-dependence in Equation~\eqref{eq2} dictates that 
only axial fundamentals are of interest here. 

The set of ideal MHD equations is evolved using the PLUTO code \citep{2007ApJS..170..228M}.
We use the piecewise parabolic method for spatial reconstruction,
and the second-order Runge–Kutta method for time-stepping.
The Harten-Lax-van Leer Discontinuities (HLLD) approximate Riemann solver 
\citep{2005JCoPh.208..315M}
is used for computing inter-cell fluxes, 
and a hyperbolic divergence cleaning method
is used to keep the magnetic field divergence-free.
We simulate only half of the loop by taking advantage of 
the symmetry with respect to the apex plane ($z=L/2$).
The simulation domain is $[-20,20]\rm{Mm}\times[-20,20]\rm{Mm}\times[0,25]\rm{Mm}$
in $x$, $y$, and $z$.
In the $x$- and $y$-direction, we use 600 uniform grids  for $[-4,4]\rm{Mm}$
and 200 stretched grids in the rest.
A uniform grid with 50  cells is adopted in the $z$-direction.
We use the outflow boundary condition for all primitive variables
at the lateral boundaries ($x$ and $y$).
Symmetric boundary condition is applied at the top boundary ($z=L/2$).
For the  bottom boundary ($z=0$),
the transverse velocities ($v_x$ and $v_y$) are fixed at zero.
The density, pressure, and $B_z$ are fixed at their initial values.
The other variables are set as outflow.

\section{Simulation Results}
\label{Results}
\begin{figure}    
	\begin{center}
		\resizebox{\hsize}{!}{\includegraphics[width=\textwidth,clip=]{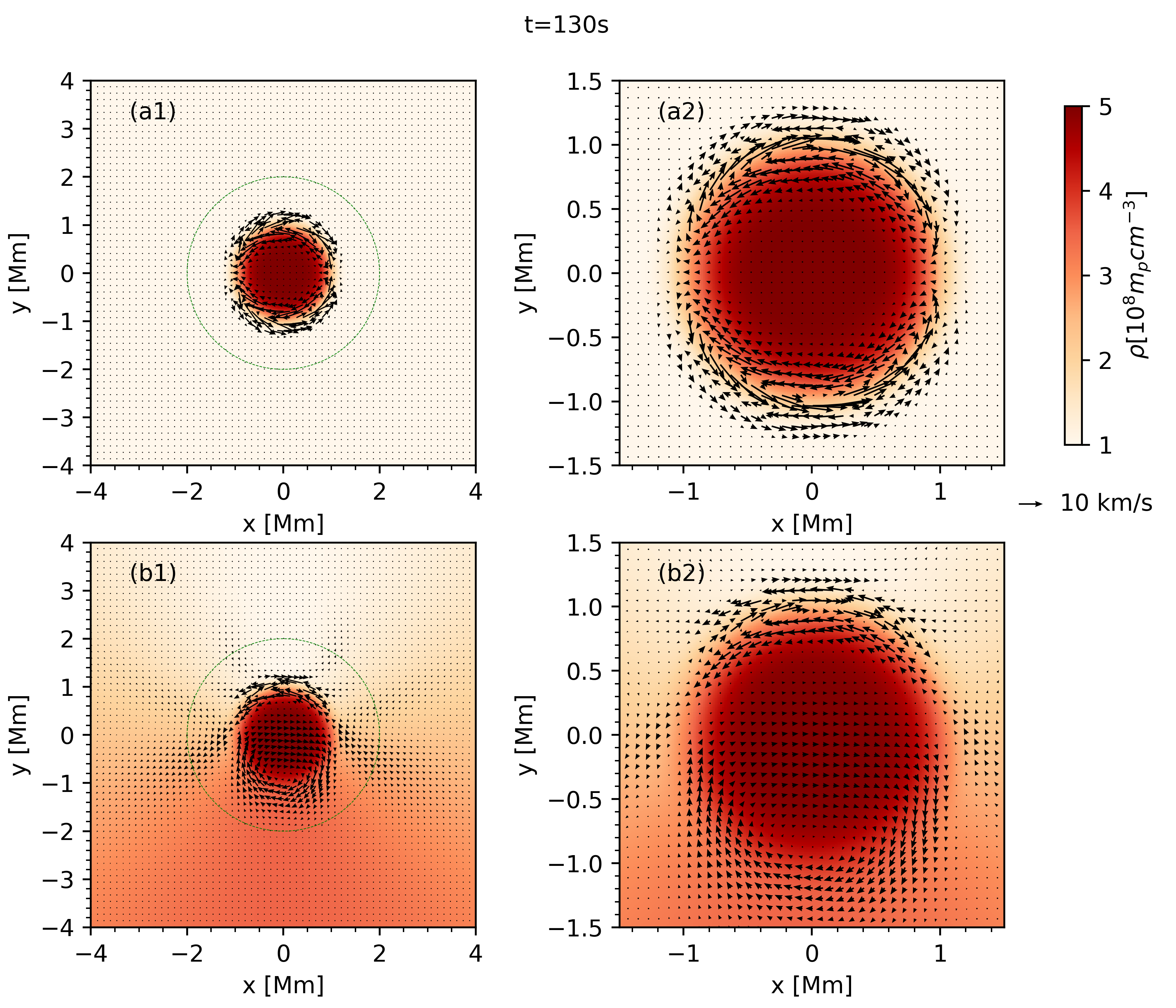}}
		\caption{Density distributions and velocity fields at the loop apex
			for the symmetric (top) and asymmetric (bottom) cases at $t=130$s.
			The green dashed lines mark a circle of radius $r=2$Mm. 
			The right panels are the zoom-in views of the left ones.
			An animated version of this figure is available that has the same layout and runs from 0 to 400 s. }
		\label{rho_vfield}
	\end{center}
\end{figure}

Figure~\ref{rho_vfield} shows the density distributions together with the velocity fields
in the apex plane at $t=130$s for the symmetric (top panels) and asymmetric (bottom) cases.
The left column displays these quantities for a large domain 
while the right column zooms in on the inner portion. 
From Figure~\ref{rho_vfield} and the associated animation,
one sees that the velocity field for the symmetric case is typical
of azimuthally standing kink motions.
The development of vortical motions is clear at the loop boundary
as a result of resonant absorption and phase mixing.
For the asymmetric case, one sees the development of fan-blade-like motions 
propagating azimuthally.
Furthermore, the velocity fields at the upper and lower loop boundaries are quite different,
with small scales much easier to tell at the upper boundary. 

\begin{figure}    
	\begin{center}
		\resizebox{\hsize}{!}{\includegraphics[width=\textwidth,clip=]{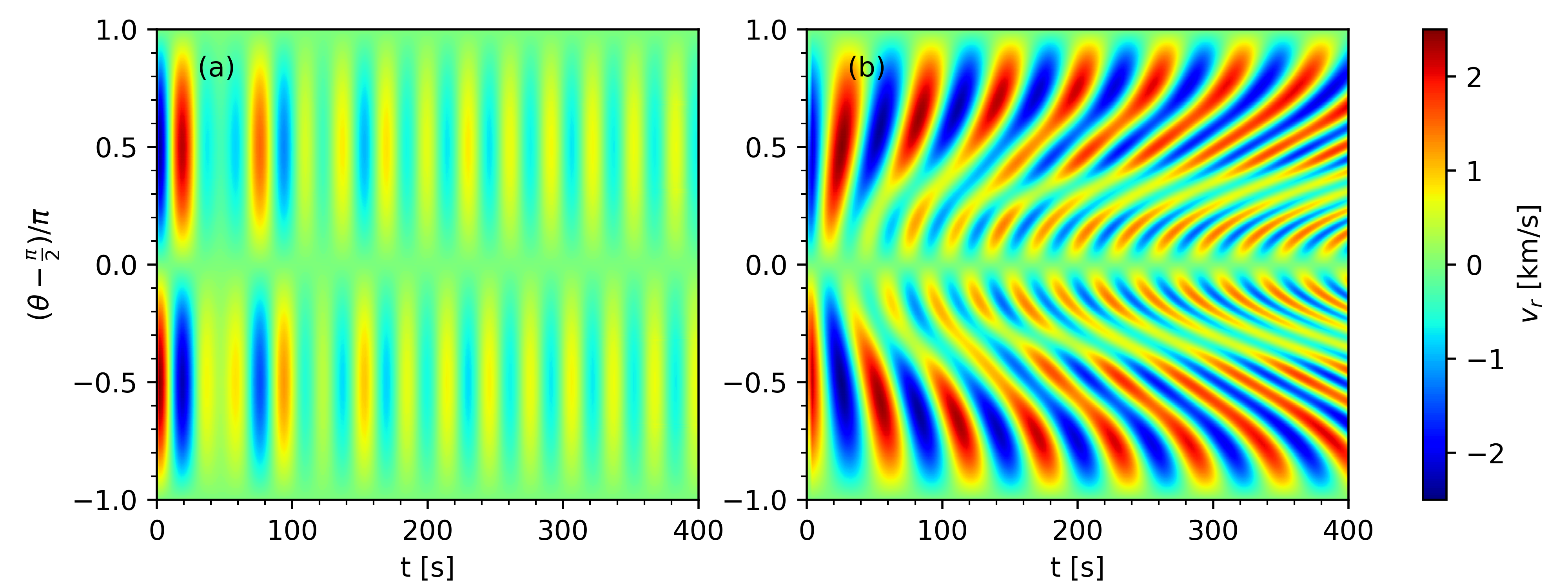}}	
		\caption{Temporal evolution of the radial velocity $v_r$ along the circle delineated in Figure~\ref{rho_vfield} for (a) the symmetric and (b) asymmetric cases.}
		\label{vr_t}
	\end{center}
\end{figure}

We further demonstrate the difference of the velocity fields between the two cases
by examining their radial velocities $v_r$.
Figure~\ref{vr_t} shows the temporal evolution of $v_r$
along a circle of radius $r=2$~{Mm} in the apex plane (the dashed lines in Figure~\ref{rho_vfield}).
For the symmetric case (a), azimuthally standing motions are clearly shown.
For the asymmetric case (b), however, $v_r$ is seen to propagate in the azimuthal direction 
as the velocity stripes become increasingly inclined with time.
Put another way, Fourier modes with increasingly high azimuthal wave numbers 
show up as time proceeds.

\begin{figure}    
	\begin{center}
		\resizebox{\hsize}{!}{\includegraphics[width=\textwidth,clip=]{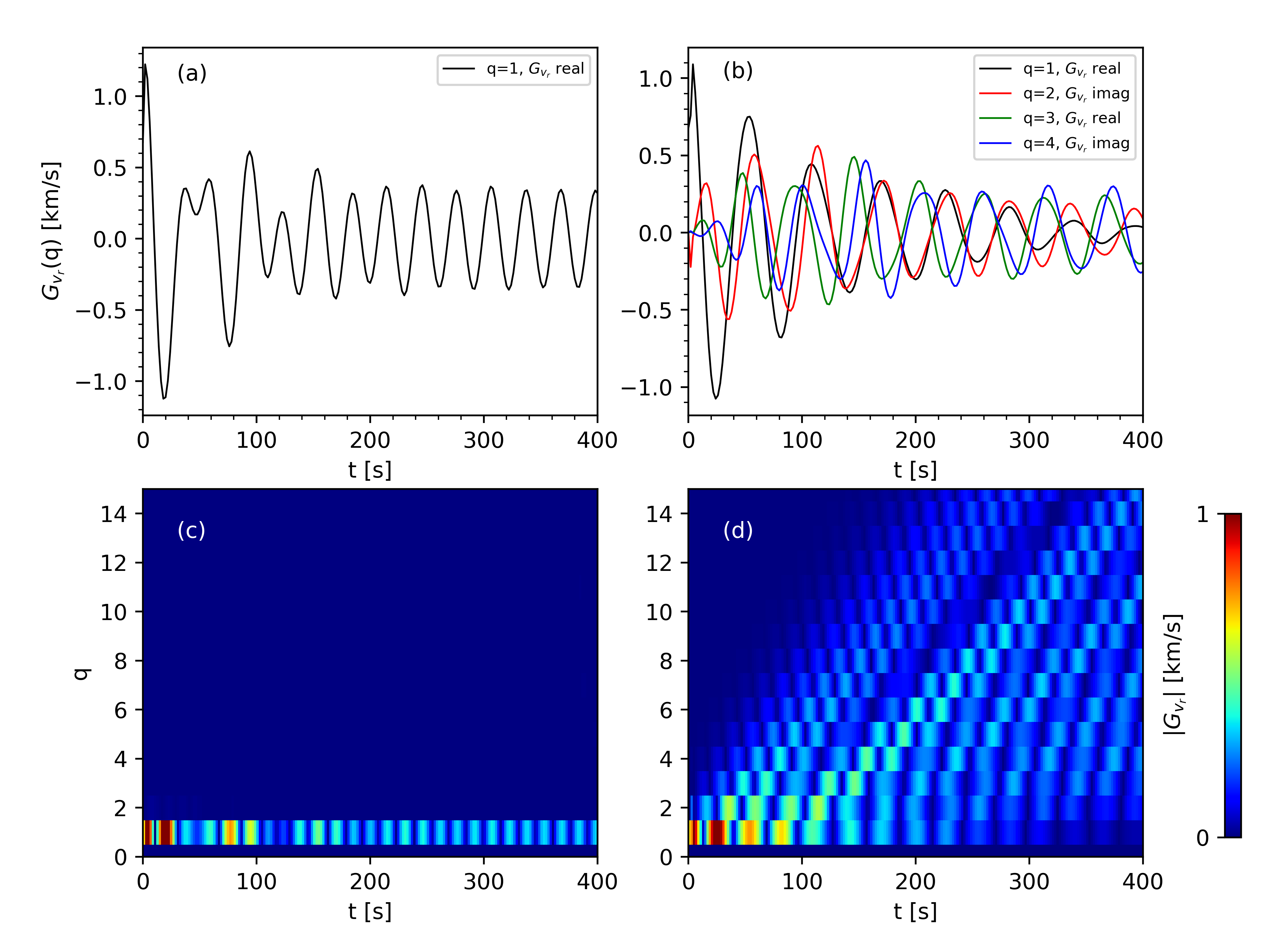}}
		\caption{Fourier coefficients $G_{v_r}$ for the symmetric (left) and asymmetric (right) cases.
			Top panels show the real or imaginary part of $G_{v_r}$ for some selected $q$ as a function of time.
			Bottom panels show the modulus $|G_{v_r}|$ as a function of time and $q$.}
		\label{Gr_t}
	\end{center}
\end{figure}

We then decompose the radial velocity $v_r$ in Figure~\ref{vr_t} into different Fourier modes.
Following \cite{2018ApJ...853...35T}, this decomposition is performed using the discrete Fourier transform, 
\begin{eqnarray}
	G_{v_r}(q) = \frac{1}{N}\sum_{n=0}^{N-1}v_r(n)e^{-i\frac{2\pi n}{N}q},
\end{eqnarray}	
where $G_{v_r}(q)$ represents the coefficient of Fourier mode $q$,
$N$ is the length of the discrete $v_r$ sequence,
and $v_r(n)$ is the $n$-th element.
Evidently, the real (imaginary) part of $G_{v_r}$ nearly vanishes for even (odd) $q$
given that $v_r$ is almost symmetric about the $y$-axis.
The Fourier coefficients as functions of $q$ and time are displayed in Figure~\ref{Gr_t}.
For the symmetric case, only a single Fourier mode of $q=1$ is recognizable,
with $G_{v_r}(q=1)$ being almost purely real (Figure~\ref{Gr_t}(a)).
For the asymmetric case, however, Fourier modes with higher $q$ are excited.
Figure~\ref{Gr_t}(b) displays the real or imaginary parts of 
$G_{v_r}(q)$ for four selected Fourier modes.
One finds that the Fourier coefficients of higher $q$ modes reach their maximum at later stages,
indicating the coupling among different Fourier modes.
The modulus $|G_{v_r}(q)|$ as a function of $q$ and time is displayed 
in the bottom panels for the (c) symmetric and (d) asymmetric cases.
Similarly, one sees a single Fourier mode of $q=1$ in the symmetric case.
For the asymmetric case, multiple Fourier modes coexist and higher $q$ modes are
successively excited as time proceeds.
These results demonstrate that the fan-blade-like velocity fields propagating in the azimuthal direction
are attributable to the continuous excitation of higher Fourier modes and the coupling among these modes.

\begin{figure}    
	\begin{center}
		\resizebox{\hsize}{!}{\includegraphics[width=\textwidth,clip=]{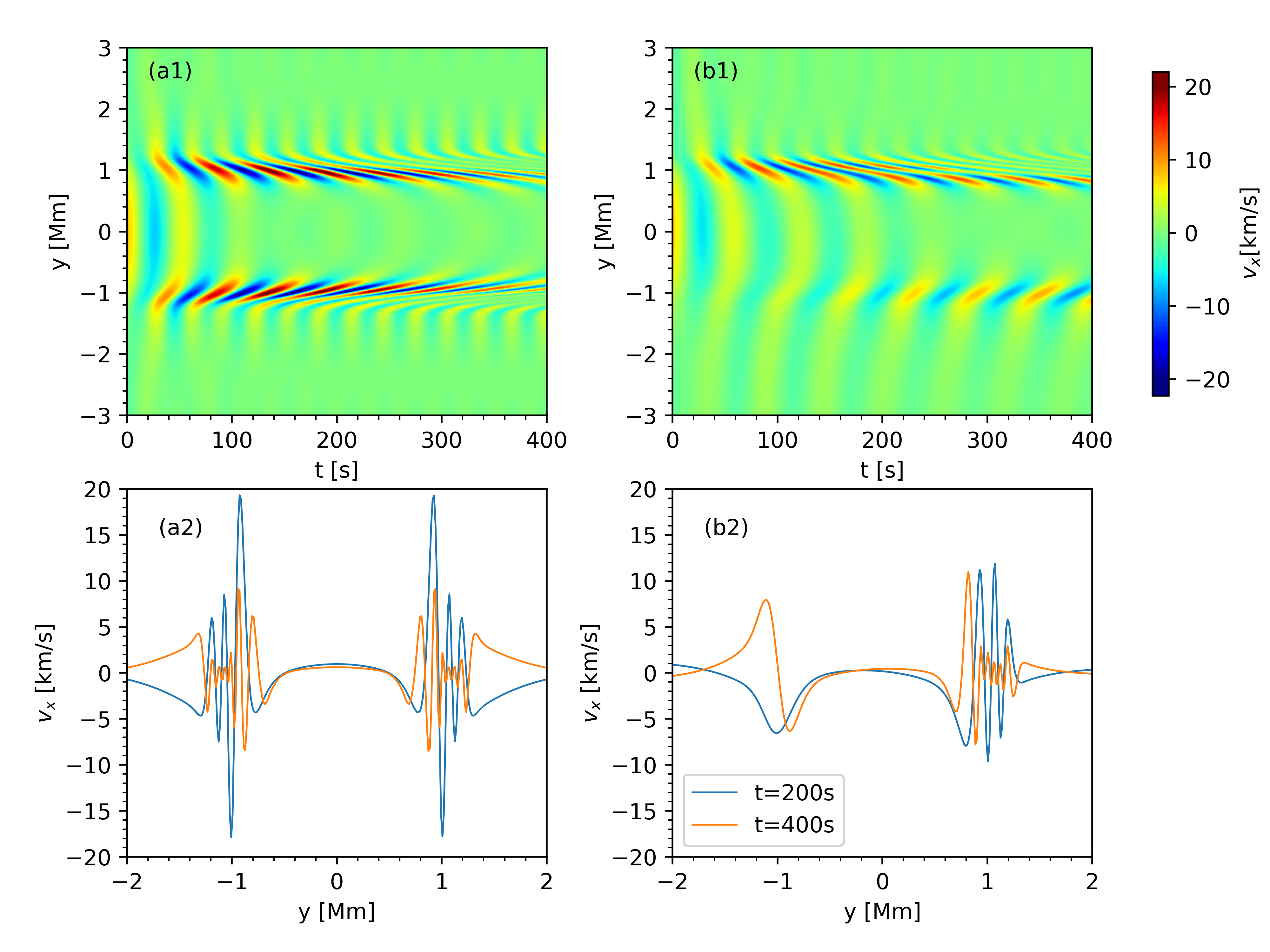}}
		\caption{Temporal evolution of $v_x(x=0,y,z=L/2)$ for (a1) the symmetric and (b1) the asymmetric cases.
			The profiles of $v_x(x=0,y,z=L/2)$ at two different times are displayed for (a2) the symmetric and (b2) asymmetric cases.
		}	
		\label{vx_y_t}
	\end{center}
\end{figure}

We have verified that wave energy is well contained in the system 
after a brief transient phase.
Kink mode is then subject to damping due to resonant absorption and the subsequent phase mixing  \citep{2011SSRv..158..289G}.
This process manifests itself as the vortical motions and the development of small scales at loop boundaries.
The top panels of Figure~\ref{vx_y_t} show the temporal evolution of $v_x$ along the $y$ axis in the apex plane ($z=L/2$).
For the symmetric case (a1), small-scale spikes are symmetrically developing at both sides of the loop boundary.
For the asymmetric case, however, spikes develop more rapidly at the upper boundary than at the lower one.
This difference is further demonstrated in the bottom panels of Figure~\ref{vx_y_t},
where two snapshots ($t=200$s and $400$s ) of $v_x$ are displayed.
We compare our results with the theoretical prediction given by \cite{1995JGR...10019441M},
from which our phase mixing length $L_{\rm ph}$ is estimated as 
\begin{eqnarray}
	L_{\rm ph} = \frac{2\pi}{|\partial\omega_{\rm A}/\partial y|}\frac{1}{t},
\end{eqnarray}
where $\omega_{\rm A}=(\pi/L)v_{\rm A}(x,y)$ is the local Alfv\'en frequency.
That small scales develop more rapidly at the upper boundary 
then derives from
the stronger gradient in the Alfv\'en frequency therein.
We proceed more quantitatively by evaluating
 the Alfv\'en frequencies
$[\omega_{\rm Ai},\omega_{\rm Ae1},\omega_{\rm Ae2}]\approx[0.09,0.21,0.11]$~rad~s$^{-1}$
with the corresponding densities. 
Examine the Alfv\'en continnum pertinent to the upper loop boundary for now.     
Equation~(20) in \citet{1995JGR...10019441M} indicates that the instantaneous
phase variation across the continuum is 
$\Delta\varphi_1=(\omega_{\rm Ae1}-\omega_{\rm Ai}) t$, 
meaning that the number of spikes is approximately 
$N_{\rm spikes, 1} \approx \Delta\varphi_1/\pi$.   
Our setup then yields that $\sim 7.6$ spikes are accumulated every $200$~secs, 
which agrees well with Figure~\ref{vx_y_t}b2. 
Likewise, one expects that $\sim 1.3$ extrema will be accumulated every $200$~secs 
close to the lower loop boundary, which is also quantitatively reproduced.

\begin{figure}    
	\begin{center}
		\resizebox{\hsize}{!}{\includegraphics[width=\textwidth,clip=]{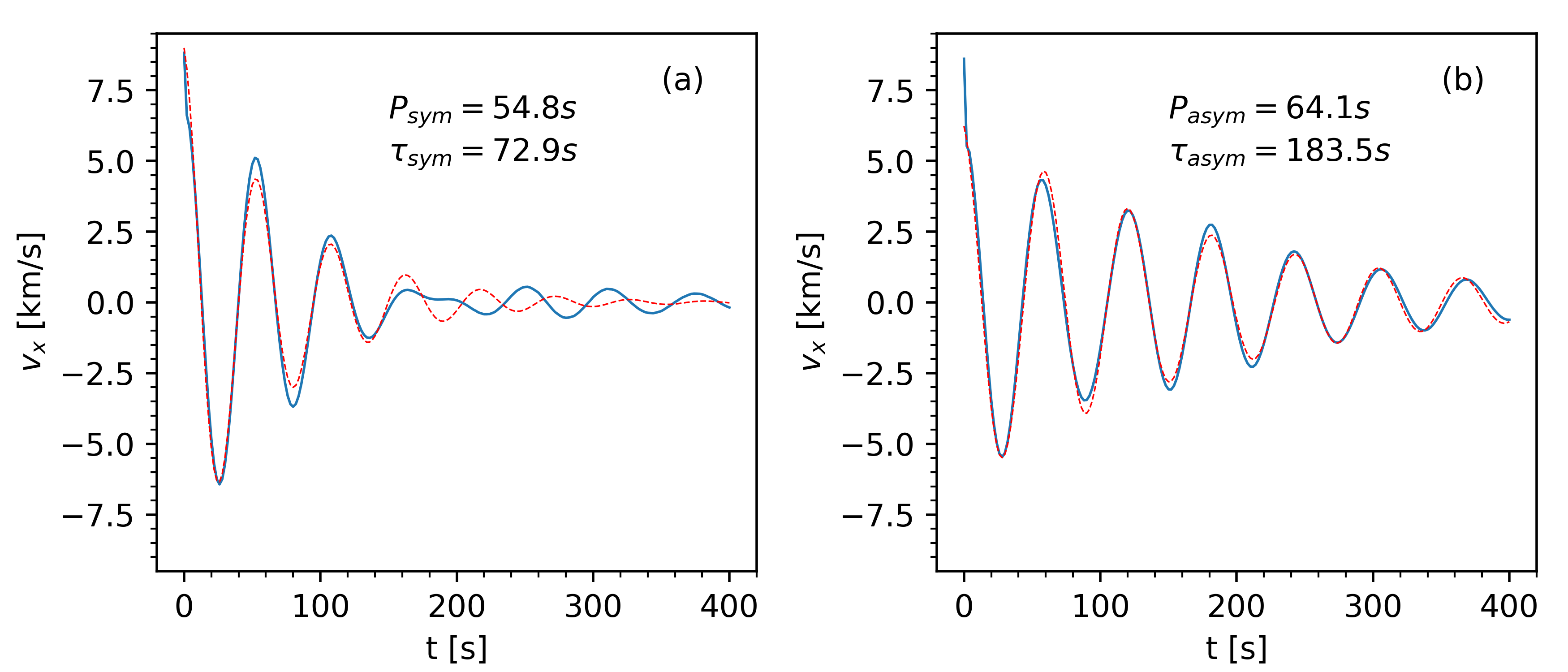}}
		\caption{Temporal profiles of $v_x$ (averaged over the loop center with $\rho>0.9\rho_i$) for (a) the symmetric and (b) the asymmetric cases.
			The red dashed lines are the best-fit exponential damped sinusoids, with the fitted period $P$ and damping time $\tau$ displayed on each panel.}
		\label{vx_damping}
	\end{center}
\end{figure}

The development of small scales is accompanied by
the damping of coordinated kink motions.
The damping envelope of kink modes has been extensively examined given their seismological potential 
\citep[e.g.,][]{2002ApJ...577..475R,2012A&A...539A..37P,2013A&A...551A..39H}.
Here we only focus on the difference of the period and damping time for the asymmetric case
from the symmetric case. 
Figure~\ref{vx_damping} displays the temporal evolution of $v_x$ averaged over the region of $\rho>0.9\rho_i$
for (a) the symmetric and (b) asymmetric cases (labeled ``sym'' and ``asym'' here).
We fit the profiles using an exponentially damped sinusoid, 
printing the best-fit periods ($P$) and damping times ($\tau$). 	
One sees that $P_{\rm asym}>P_{\rm sym}$, which is unsurprising given an overall heavier 
ambient corona and hence a larger inertia experienced by the kink motions
in the asymmetric case.
That $\tau_{\rm asym}>\tau_{\rm sym}$ can be understood by drawing analogy with
the damping of kink motions in an axisymmetric setup, for which the ideal quasi-mode theory predicts that $\tau/P$ in the thin-boundary-thin-tube (TTTB) limit
decreases with the ratio of the internal density
$\rho_{\rm i}$ to the ambient value $\rho_{\rm amb}$
\citep[e.g.,][]{1992SoPh..138..233G,2002ApJ...577..475R}.
While not straightforward to quantify, $\rho_{\rm amb}$ in the asymmetric case 
is expected to exceed
$\rho_{\rm e1}$, which can be unambiguously identified as $\rho_{\rm amb}$
in the symmetric case. 
That said, an exponential damping envelope does not apply to the symmetric case
at large times (Figure~\ref{vx_damping}a), reinforcing the notion that ideal quasi-modes 
offer only a partial picture for the time-dependent coordinated motions \citep[see][for details]{2015ApJ...803...43S}.

The profile of $v_x$ in the symmetric case (Figure~\ref{vx_damping}a) shows a beat at $t\approx160{\rm s}$.
One  possible cause of this beat is the interplay between kink motions and  Alfv\'enic motions \citep[e.g.,][Fig.9]{2015ApJ...803...43S}.
It is known that the presence of similar beats may lead to an underestimation of the actual damping time \citep{2013A&A...552A..57N}.
We therefore also compute the period ($P_{\rm QM}$) and damping time ($\tau_{\rm QM}$) expected for kink quasi-modes (QMs) resonantly absorbed in the Alfv\'en continuum \citep[see the review by][for conceptual clarifications]{2011SSRv..158..289G}. 
We specificially adapt our previous resistive eigenmode code \citep{2021ApJ...908..230C} to address the present equilibrium configuration, 
finding that  $P_{\rm QM}\approx 51.6~{\rm sec}$ and $\tau_{\rm QM} \approx 67.1~{\rm sec}$. 
These QM values are rather close to the best-fit values in Figure~\ref{vx_damping}(a)($P_{\rm sym}$ and $\tau_{\rm sym}$), suggesting that the presence of a beat-like behavior may not significantly affect our fitting approach for evaluating the kink periods and damping times.

\begin{figure}    
	\begin{center}
		\resizebox{\hsize}{!}{\includegraphics[width=\textwidth,clip=]{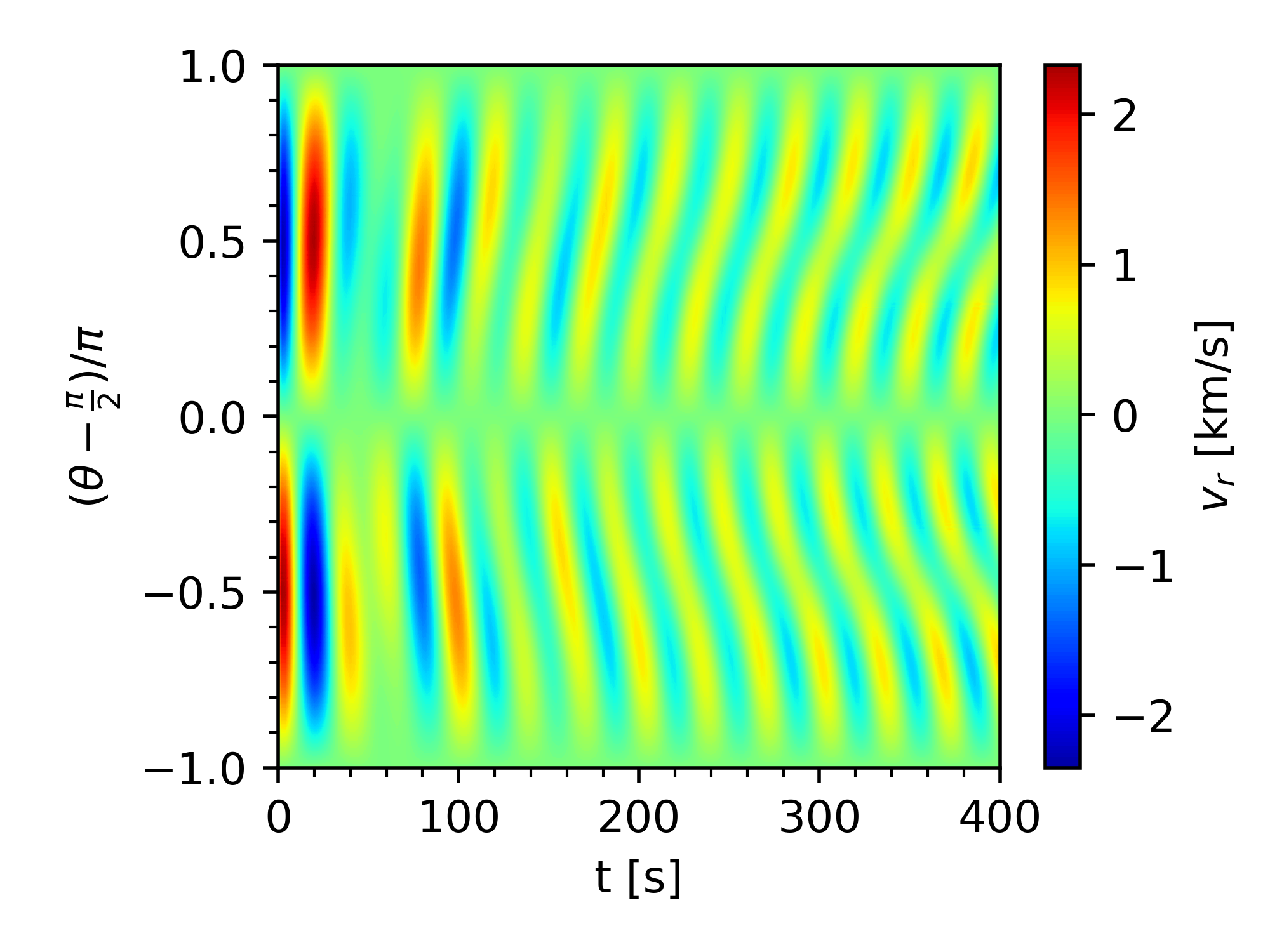}}
		\caption{Similar to Figure~\ref{vr_t}b, but for the case of $\rho_{\rm e2}=1.2$.}
		\label{vr_t_V1}
	\end{center}
\end{figure}

\begin{figure}    
	\begin{center}
		\resizebox{\hsize}{!}{\includegraphics[width=\textwidth,clip=]{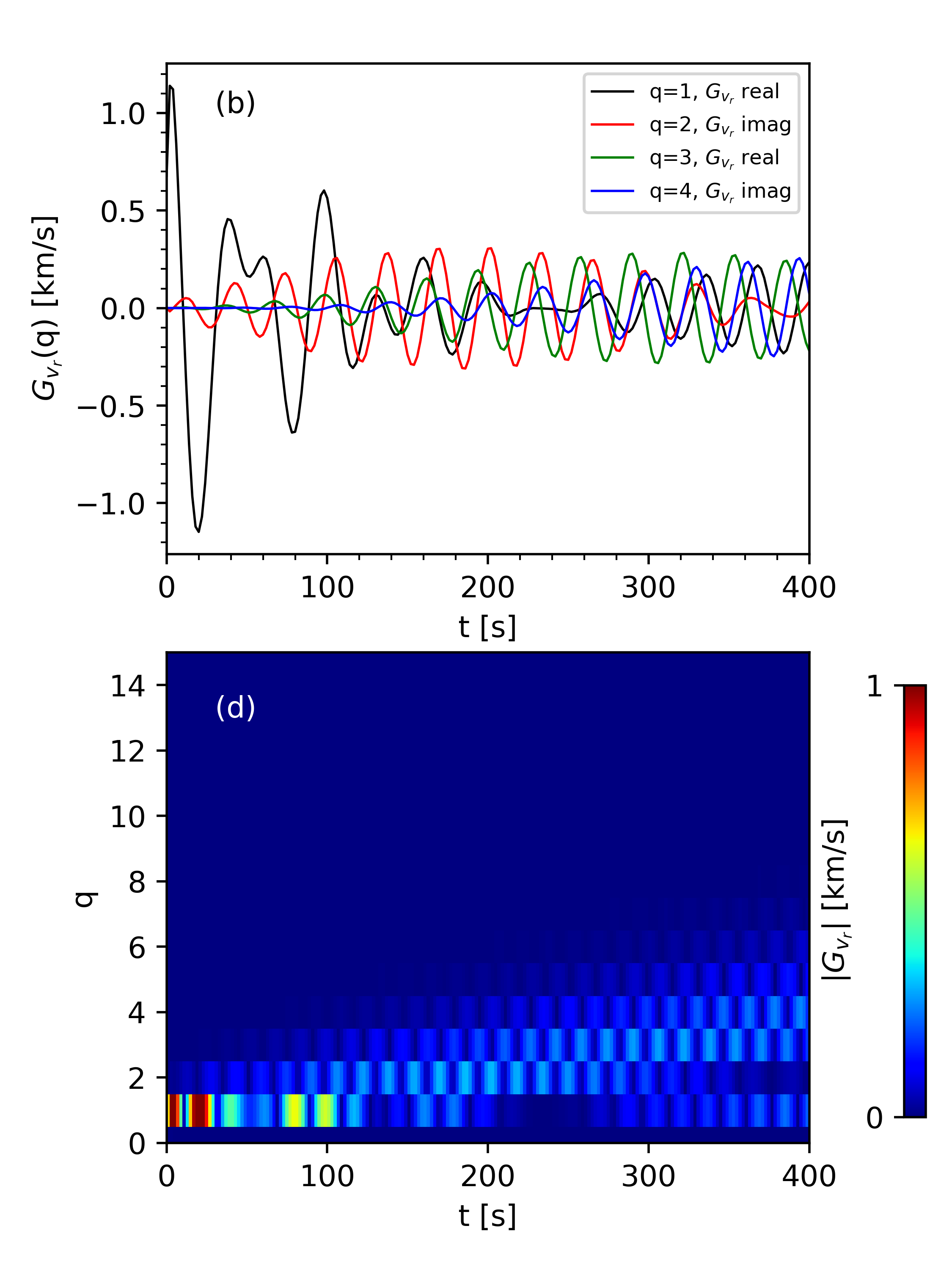}}
		\caption{Similar to the right column of Figure~\ref{Gr_t}, but for the case of $\rho_{\rm e2}=1.2$.}
		\label{Gr_t_V1}
	\end{center}
\end{figure}

In this work, we pay attention to the effect of  an asymmetric loop environment on kink oscillations,
and thus neglect both gravity and the gravitationally stratified atmosphere in order to avoid further complexity.
The density profile we adopted (Equation~\ref{rho_profile}) may not quite well represent the realistic stratified density profile.
In terms of modeling, however, our model makes a difference with most previous simulations where a uniform background was assumed.
The results shown here are relevant for further exploring more realistic models.

It is necessary to mention that for the asymmetric case, the density variation of the environment ($\rho_{\rm e2}/\rho_{\rm e1}=3.5$) 
across the loop width is more significant  than
the density variations over a scale height ($\Lambda\approx50{\rm Mm}$) of a $1$MK corona.
While we do not attempt to model a stratified density profile,
we nonetheless perform an additional simulation with $\rho_{\rm e2}=1.2$.
This density ratio of $\rho_{\rm e2}/\rho_{\rm e1}=1.2$ is in line with some 
loops observed by TRACE or SOHO/EIT whose width $w$ can be as large as 8Mm \citep[e.g.,][]{2007ApJ...662L.119S},
i.e., ${\rm exp}(w/\Lambda)\approx1.2$
\footnote{Note that a nominal loop radius $R=1~{\rm Mm}$ serves essentially as a normalizing constant in our gravity-free modeling. The key dynamics here results primarily from the deviation of the dimensionless $\rho_{\rm e2}/\rho_{\rm e1}$ from unity. As such, the ensuing results are equally applicable upon proper scaling \citep[see Sect.4.1.2 in][for more on this aspect]{2019CUP_goedbloed_keppens_poedts}.}.
Similar analyses are performed, with the results being shown in Figure~\ref{vr_t_V1} and Figure~\ref{Gr_t_V1}.
We find that the key features we previously obtained, the azimuthally propagating velocity field and the excitation of higher azimuthal Fourier modes, still exist, 
though not as strong as the case with $\rho_{\rm e2}=3.5$. 
Our model thus provides a proof-of-concept test of the effect of an asymmetric environment on kink oscillations.
More realistic models taking into account both the stratified atmosphere and the gravity effect are called for.

\section{Summary}
\label{S-summary}
We studied kink modes supported by a straight coronal loop
in an asymmetric environment using 3D MHD simulations.
The new features compared with the case of symmetric ambients are summarized as follows:
\begin{itemize}
	\item[1)] The velocity fields in the exterior shows the development of fan-blade features
	that propagate azimuthally.
	This is attributed to the successive excitation of higher azimuthal Fourier modes.
	\item[2)] Resonant absorption and phase mixing can still occur in the asymmetric case.
	Velocity spikes are generated nonsymmetrically,
	with small scales developing more rapidly at the upper boundary than at the lower.
\end{itemize}

This work provides a simplified model studying horizontally polarized kink modes
supported by coronal loops in an asymmetric ambient corona. 
These findings enrich our understanding of kink modes in more realistic configurations
and are helpful for future high-resolution observations.

\begin{acknowledgements}
	This work is supported by the National Natural Science Foundation of China (12273019, 41974200, 42230203, 41904150).
	We gratefully acknowledge ISSI-BJ for supporting the international team “Magnetohydrodynamic wavetrains as a tool for probing the solar corona ”.
\end{acknowledgements}

\bibliographystyle{aa} 
\bibliography{export-bibtex} 

\end{document}